\documentclass[twocolumn,showpacs,preprintnumbers,amsmath,amssymb,floatfix,nofootinbib,a4]{revtex4}

\usepackage{amsmath,amsfonts,bbm,euscript,hyperref}
\usepackage{graphicx}
\usepackage{dcolumn}
\usepackage{bm}
\usepackage{epsfig}
\epsfclipon

\newcommand{\nco}{\newcommand}
\nco{\beq}{\begin{equation}} \nco{\eeq}{\end{equation}}
\nco{\beqa}{\begin{eqnarray}} \nco{\eeqa}{\end{eqnarray}}
\def\be{\begin{equation}}
\def\ee{\end{equation}}    
\def\baray{\begin{eqnarray}}
\def\earay{\end{eqnarray}}

\nco{\lra}{\leftrightarrow}

\nco{\sss}{\scriptscriptstyle} \nco{\dphi}{\varphi}
\nco{\lsim}{\mbox{\raisebox{-.6ex}{~$\stackrel{<}{\sim}$~}}}
\nco{\gsim}{\mbox{\raisebox{-.6ex}{~$\stackrel{>}{\sim}$~}}}

\def\IK{\relax{\rm I\kern-.20em K}}
\def\IM{\relax{\rm I\kern-.20em M}}

\def\lsim{\mbox{\raisebox{-.6ex}{~$\stackrel{<}{\sim}$~}}}
\def\gsim{\mbox{\raisebox{-.6ex}{~$\stackrel{>}{\sim}$~}}}
\def\sss{\scriptscriptstyle}

\begin{document}

\title{Particle Production During Inflation: Observational Constraints and Signatures}

\author{Neil Barnaby$^{1}$, Zhiqi Huang$^{2}$}

\affiliation{%
\centerline{Canadian Institute for Theoretical Astrophysics,}
\centerline{University of Toronto, McLennan Physical Laboratories,
60 St.\ George Street, Toronto, Ontario, Canada  M5S 3H8}
${}^1$e-mail:\ barnaby@cita.utoronto.ca, ${}^2$e-mail:\ zqhuang@astro.utoronto.ca}

\begin{abstract} 
In a variety of inflation models the motion of the inflaton may trigger the production of some non-inflaton particles during inflation, for
example via parametric resonance or a phase transition.  Particle production during inflation leads to observables in the cosmological fluctuations,
such as features in the primordial power spectrum and also nongaussianities.  Here we focus on a prototype scenario with  inflaton, $\phi$,
and iso-inflaton, $\chi$, fields interacting during inflation via the coupling $g^2 (\phi-\phi_0)^2\chi^2$.  Since several previous investigations have hinted at the presence
of localized ``glitches'' in the observed primordial power spectrum, which are inconsistent with the simplest power-law model, it is interesting to determine the extent
to which such anomalies can be explained by this simple and microscopically well-motivated inflation model.  Our prototype scenario predicts a 
bump-like feature in the primordial power spectrum, rather than an oscillatory ``ringing'' pattern as has previously been assumed.
We discuss the observational constraints on such features using a variety of cosmological data sets.
We find that bumps with amplitude as large as $\mathcal{O}(10\%)$ of the usual scale invariant fluctuations from inflation, corresponding to $g^2 \sim 0.01$,
are allowed on scales relevant for Cosmic Microwave Background experiments.  Our results imply an upper 
limit on the coupling $g^2$ (for a given $\phi_0$) which is crucial for assessing the detectability of the nongaussianity produced by inflationary particle production.  
We also discuss more complicated features
that result from superposing multiple instances of particle production.  Finally, we point to a number of microscopic 
realizations of this scenario in
string theory and supersymmetry and discuss the implications of our constraints for the popular brane/axion monodromy inflation models.
\end{abstract}

\maketitle

\section{Introduction}

Recently, there has been considerable interest in inflationary models where the motion of the inflaton triggers the production of
some non-inflation (iso-curvature) particles \emph{during} inflation 
\cite{chung,chung2,elgaroy,sasaki,ir,modulated,KL,KP,BBS,ng,adams,step_model,gobump,beauty,trapped,newref,warm}.  
Examples have been studied where this particle production
occurs via parametric resonance \cite{chung,chung2,elgaroy,sasaki,ir,modulated}, as a result of a phase transition 
\cite{KL,KP,BBS,ng,adams,step_model,gobump}, or otherwise.  
In some scenarios, backreaction effects from particle production can slow the motion of the inflaton on a steep potential \cite{beauty,trapped,newref}, providing a new inflationary mechanism.  Moreover, inflationary
particle production arises naturally in a number of realistic microscopic models from string theory \cite{beauty,trapped,newref,monodromy,monodromy2,monodromy3} and also supersymmetry (SUSY) \cite{berrera}.

In \cite{ir} it was shown that the production of massive iso-curvature particles during inflation (and their subsequent interactions with the slow roll condensate) provides a qualitatively new
mechanism for generating cosmological perturbations.  This new mechanism leads to a variety of novel observable signatures, such as features \cite{ir} and nongaussianities \cite{ir,inprog2} in the
primordial fluctuations.  In this paper we study in detail the observational constraints on such distortions of the primordial power spectrum for a variety of scenarios.

One motivation for this study is to determine whether features generated by particle production during inflation can explain some of the 
anomalies in the observed primordial 
power spectrum, $P(k)$.  A number of different studies have hinted at the possible presence of some localized features 
in the power spectrum \cite{chung2,gobump,features,features2,morefeatures1,morefeatures2,morefeatures3,features3,yokoyama1,yokoyama3,yokoyama2,hoi1,hoi,contaldi}, which are not compatible with the simplest power law $P(k) \sim k^{n_s - 1}$
model.  Although such glitches may simply be statistical anomalies, there is also
the tantalizing possibility that they represent a signature of primordial physics beyond the simplest slow roll inflation scenario.  Forthcoming
polarization data may play a crucial role in distinguishing between these possibilities \cite{gobump}.  However, in the meantime, it is interesting to determine
the extent to which such features may be explained by microscopically realistic inflation models.

We consider a very simple model where the inflaton, $\phi$, and iso-inflaton, $\chi$, fields interact via the coupling
\begin{equation}
\label{int}
\mathcal{L}_{\mathrm{int}} = -\frac{g^2}{2}(\phi-\phi_0)^2 \chi^2
\end{equation}
We focus on this simple prototype model in order to illustrate the basic phenomenology of particle production during inflation, however,
we expect our results to generalize in a straightforward way to more complicated scenarios. Models of the type (\ref{int}) have been considered as a 
probe of Planck-scale effects \cite{chung} and offer a novel example of the non-decoupling of high energy physics during inflation.\footnote{For reasonable values of $g^2$ the $\chi$
particles are \emph{extremely} massive for almost the entire duration of inflation excepting a tiny interval, much less than an $e$-folding,
about the point $\phi=\phi_0$.  However, the $\chi$ field cannot be integrated
out due to the non-adiabatic time dependence of the mode functions, see \cite{jim} for further discussion.}

At the moment when $\phi=\phi_0$ (which we assume occurs during the observable range of $e$-foldings of inflation) the $\chi$ particles become 
instantaneously  massless and are produced by quantum effects.  
This burst of particle production  drains energy from the condensate $\phi(t)$, temporarily slowing the 
motion of the inflaton background and violating slow roll.  Shortly after this moment the 
$\chi$ particles become extremely non-relativistic, so that their number density dilutes as $a^{-3}$, and eventually the inflaton 
settles back onto the slow roll attractor. 

Several previous papers \cite{chung,chung2,elgaroy,sasaki} have studied the temporary slowing-down of the inflaton
background using the mean-field equation
\begin{equation}
\label{mean}
  \ddot{\phi} + 3H \dot{\phi} + V_{,\phi} + g^2 (\phi-\phi_0) \langle \chi^2 \rangle = 0
\end{equation}
where the vacuum average $\langle \chi^2 \rangle$ is computed following \cite{kls1,kls2}.  Using this approach one 
finds that the transient violation of slow roll leads to a ``ringing pattern'' (damped oscillations) in the spectrum 
of cosmological fluctuations leaving the horizon near the moment when $\phi=\phi_0$ \cite{sasaki}.  In \cite{chung2,elgaroy} observational
constraints on particle production during inflation were discussed in the context of this mean field treatment.

\begin{figure}[htbp]
\bigskip \centerline{\epsfxsize=0.15\textwidth\epsfbox{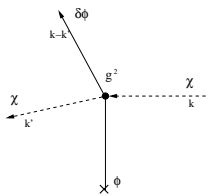}}
\caption{Rescattering diagram.
}
\label{Fig:diag}
\end{figure}

\begin{figure}[htbp]
\bigskip \centerline{\epsfxsize=0.35\textwidth\epsfbox{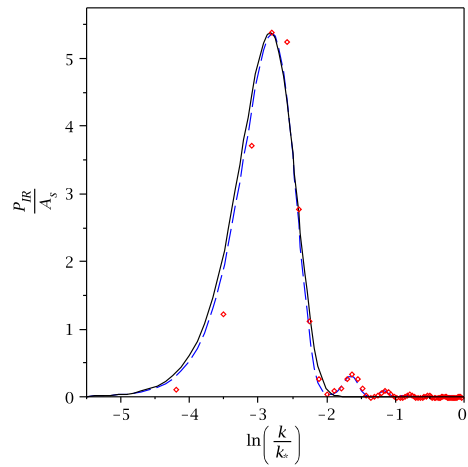}}
\caption{The bump-like features generated by IR cascading.  We plot the feature power spectrum obtained from fully nonlinear lattice field theory
      simulations (the red points) and also the result of an analytical calculation (the dashed blue curve) using the formalism described in
      \cite{inprog1}.  We also superpose the fitting function $\sim k^3 e^{-\pi k^2 / (2 k_\star^2)}$ (the solid black curve) to illustrate the accuracy
      of this simple formula
}
\label{Fig:BumpFit}
\end{figure}

However, in \cite{ir} cosmological fluctuations in the model (\ref{int}) were
re-considered, going beyond the mean-field treatment of $\phi$.\footnote{See also \cite{trapped} for a complimentary
analysis and \cite{inprog1} for a detailed analytical treatment of the dynamics.}  It was pointed out 
that the massive $\chi$ particles can rescatter off the condensate to generate 
bremsstrahlung radiation of long-wavelength $\delta \phi$  fluctuations via diagrams such as Fig.~\ref{Fig:diag}.  
Multiple rescattering processes lead to a rapid cascade of power into the infra-red (IR) - \emph{IR cascading}.  The 
inflaton modes generated by IR cascading freeze once their wavelength crosses the horizon and lead to a bump-like 
feature in the cosmological perturbations that is illustrated in Fig.~\ref{Fig:BumpFit}.  This feature is complimentary to the usual (nearly) 
scale-invariant quantum vacuum 
fluctuations from inflation.  The bump dominates over the ringing pattern discussed above by many orders of magnitude,
independently of the value of $g^2$.  

In light of the results of \cite{ir} it is clear that the observational constraints on the model (\ref{int}) need to be reconsidered.  
Since previous studies have suggested marginal evidence for localized power excesses in the CMB using both parametric \cite{chung2,hoi}
and non-parametric \cite{yokoyama2,yokoyama3} techniques, it is interesting to determine if a simple and well-motivated model such as (\ref{int})
can explain these anomalies.  To answer this question we provide a simple semi-analytic
fitting function that accurately captures the shape of the feature generated by particle production and IR cascading
during inflation.  Next, we confront this modified power spectrum with a variety
of observational data sets.  We find no evidence for a detection, however, we note that observations are consistent with
relatively large spectral distortions of the type predicted by the model (\ref{int}).  If the feature is located on scales relevant for Cosmic Microwave Background
(CMB) experiments then its amplitude may be as large as $\mathcal{O}(10\%)$ of the usual scale-invariant fluctuations, corresponding to $g^2 \sim 0.01$.
Our results translate into a $\phi_0$-dependent bound on the coupling $g^2$ which is crucial in order to determine whether the nongaussian signal associated with particle production and IR 
cascading is detectable in future missions \cite{inprog2}.

We also consider the more complicated features which result from multiple bursts of particle production and IR cascading.  Such features are
a prediction of a number of string theory inflation models, including brane/axion monodromy \cite{monodromy,monodromy2,monodromy3}. 
For appropriate choice of the spacing between the features, we find that the constraint on $g^2$ in this scenario is even weaker than the single-bump
case.

Although we focus on the interaction (\ref{int}), our findings may have some bearing also on models 
with phase transitions during inflation.
A simple prototype model for the latter scenario is
\[
  V(\phi,\chi) = \frac{\lambda}{4}(\chi^2-v^2)^2 + \frac{g^2}{2}\chi^2\phi^2 + V_{\mathrm{inf}}(\phi)
\]
At the moment when $\phi = \sqrt{\lambda} v / g$
massive iso-inflaton $\chi$ particles are produced copiously by tachyonic (spinodal) instability \cite{tac}.  These produced particles will subsequently 
interact
with the condensate via rescattering diagrams similar to Fig.~\ref{Fig:diag}.  Hence, we expect the features produced by inflationary
phase transitions to be qualitatively similar to the bumps considered in this paper.  (This intuition is consistent with a second order computation
of the cosmological perturbations in a closely related model \cite{ng}.  See \cite{hoi} for a discussion of the observational consequences.)

In the literature it is sometimes argued that inflationary phase transitions can be studied using a toy model with a sharp step-like
feature in the inflaton potential.   This potential-step model predicts a ringing pattern in the power spectrum,
very much analogous to the mean field treatment of resonant particle 
production during inflation, discussed above.  This treatment does not take into account
the violent growth of inhomogeneities of the fields that occurs during realistic phase transitions \cite{tac} and, in particular, does not capture
rescattering effects and IR cascading.  In the case of resonant particle production, these nonlinear effects have a \emph{huge} impact on the cosmological
fluctuations \cite{ir}.  Hence, it is far from clear if the potential-step model provides a good effective description of inflationary phase transitions.\footnote{See also \cite{space_break} for a related discussion.}

Of course, inflation models with steps in $V(\phi)$ (or its derivatives) may be considered on phenomenological
grounds, irrespective of the motivation from inflationary phase transitions.  In \cite{star1,star2}
cosmological perturbations from models with step-like features and discontinuities in higher derivatives were considered, as were
the microscopic motivations for such constructions.  See \cite{chen1,chen2} for a study of the
nongaussianities induced in a variety of single-field models with steps or oscillations in the inflaton potential.

The outline of this paper is as follows.  In section \ref{sec:param} we provide a simple parametrization of the features that are
imprinted on the primordial power spectrum by one or more bursts of particle production during inflation.  In section \ref{sec:method} we describe 
our method and discuss the observational data sets employed to derive constraints on this modified power spectrum.  
In section \ref{sec:cons} we present observational constraints on various scenarios.
In section \ref{sec:particle} we present some microscopic realizations of our scenario and discuss the implications of our findings for
popular string theory/SUSY inflation models with a special emphasis on brane monodromy.  Finally, in section \ref{sec:concl} we conclude.

\section{A Simple Parametrization of the Power Spectrum\label{sec:param}}

In \cite{ir} it was shown that particle production and IR cascading during inflation
in the model (\ref{int}) generates a bump-like contribution to the primordial
power spectrum.  As shown in Fig.~\ref{Fig:BumpFit}, this feature can be fit with a very simple function
$P_{\mathrm{bump}} \sim k^3 e^{-\pi k^2 / (2 k_\star^2)}$.
The bump-like contribution from IR cascading is complimentary to the usual
(nearly) scale-invariant contribution to the primordial power spectrum $P_{\mathrm{vac}} \sim k^{n_s-1}$ coming from
the quantum vacuum fluctuations of the inflaton.  The total, observable, power spectrum in the model
(\ref{int}) is simply the superposition of these two contributions: $P(k) \sim k^{n_s-1} +  k^3 e^{-\pi k^2 / (2 k_\star^2)}$.
This simple formula can be motivated
from analytical considerations \cite{ir,inprog1} and provides a good fit to lattice field theory simulations near the peak of the feature
and also in the IR tail.\footnote{This fitting formula does \emph{not} capture the small oscillatory structure in the UV tail of the feature (see Fig.~\ref{Fig:BumpFit}) which 
does not concern us since that region is not phenomenologically interesting.}

It is straightforward to generalize this discussion to allow for multiple bursts of particle production during inflation.  
Suppose there are multiple points $\phi=\phi_i$ ($i=1,\cdots,n$) along the 
inflationary trajectory where new degrees of 
freedom $\chi_i$ become massless:
\begin{equation}
\label{multiple}
  \mathcal{L}_{\mathrm{int}} = -\sum_{i=0}^{n} \frac{g_i^2}{2}(\phi-\phi_i) \chi_i^2
\end{equation}
For each instant $t_i$ when $\phi=\phi_i$ there will be an associated burst of particle production  and subsequent rescattering of
the produced massive $\chi_i$ off the condensate $\phi(t)$.  Each of these events proceeds as described above and leads to a new
bump-like contribution to the power spectrum.  
These features simply superpose owing to that fact that each field $\chi_i$ is statistically independent (so that the cross terms involving
$\chi_i\chi_j$ with $i\not= j$ in the computation of the two-point function must vanish).  Thus, we arrive at the following parametrization of the
primordial power spectrum in models with particle production during inflation:
\begin{eqnarray}
  P(k) &=& A_s \left(\frac{k}{k_0}\right)^{n_s-1} + \nonumber \\
                    &&\,\,\,\,\,\,\sum_{i=1}^n A_i \left(\frac{\pi e}{3}\right)^{3/2}\left(\frac{k}{k_i}\right)^3 e^{-\frac{\pi}{2}\left(\frac{k}{k_i}\right)^2}    \label{param}
\end{eqnarray}
where $A_s$ is the amplitude of the usual nearly scale invariant vacuum fluctuations from inflation and $k_0$ is the pivot, which we choose
to be $k_0 = 0.002 \,\mathrm{Mpc}^{-1}$ following \cite{Hinshaw2008}.  The constants $A_i$ depend on the couplings $g_i^2$ and measure the size of the
features from particle production.  We have normalized these amplitudes so that the power in the $i$-th bump, measured at the
peak of the feature, is given by $A_i$.
The location of each feature, $k_i$, is related to the number of $e$-foldings $N$ from the end of inflation to the time when the $i$-th burst of particle
production occurs: roughly $\ln( k_i / H) \sim N_i$ where $N=N_i$ at the moment when $\phi=\phi_i$.  From a purely
phenomenological perspective the locations $k_i$ are completely arbitrary.

We compare (\ref{param}) to lattice field theory simulations in order to determine the amplitude $A_i$ in terms of model parameters.  We find
\begin{equation}
\label{Ag}
  A_i \cong 1.01\times 10^{-6}\, g_i^{15/4}
\end{equation}
Assuming standard chaotic inflation $V = m^2\phi^2 / 2$ we have tested this formula for $g^2 = 1, 0.1, 0.01$, taking both $\phi_0 = 2 M_p$ and $\phi_0 = 3.2 M_p$.
We found agreement up to factors order unity in all cases.

Theoretical consistency of our calculation of the shape of the feature bounds the coupling as $10^{-7} \lsim g^2_i \lsim 1$ \cite{ir}.
Hence, the power spectrum (\ref{param}) can be obtained from sensible microphysics only when $10^{-20} \lsim A_i \lsim 10^{-6}$.
This constraint still allows for a huge range of observational possibilities: near the upper bound the feature is considerably larger
than the vacuum fluctuations while near the lower bound the feature is completely undetectable.

Note that for each bump in (\ref{param}) the IR tail $P_{\mathrm{bump}} \rightarrow k^3$ as $k\rightarrow 0$ is similar to the feature considered by Hoi, Cline \& Holder in \cite{hoi},
consistent with causality arguments about the generation of curvature perturbations by local physics.

\section{Data Sets and Analysis\label{sec:method}}

The primordial power spectrum for our model is parametrized as (\ref{param}).  Our aim is to derive observational constraints on the various
model parameters $A_s$, $n_s$, $k_i$ and $A_i$ using CMB, galaxy power spectrum and weak lensing data.  To this end we use the cosmoMC
package \cite{Lewis2002} to run Markov Chain Monte Carlo (MCMC) calculations to determine the likelihood of the cosmological 
parameters, including our new parameters $A_{i}$ and $k_{i}$.  We employ the following data sets.

\begin{description}
\item{\it Cosmic Microwave Background (CMB)}

Our complete CMB data sets include WMAP-5yr \cite{Komatsu2008,Hinshaw2008}, BOOMERANG \cite{Jones2006,Piacentini2006,Montroy2006}, 
ACBAR \cite{Runyan2003, Goldstein2003,Kuo2006,Reichardt2008}, CBI \cite{Pearson2003,Readhead2004a,Readhead2004b,Sievers2007}, 
VSA \cite{Dickinson2004}, DASI \cite{Halverson2002,Leitch2005}, and MAXIMA \cite{Hanany2000}. We have included the Sunyaev-Zeldovic (SZ) 
secondary anisotropy \cite{Sunyaev1972,Sunyaev1980} for WMAP-5yr, ACBAR and CBI data sets. The SZ template is obtained from hydrodynamical 
simulation \cite{Bond2005}. Also included for theoretical calculation of CMB power spectra is the CMB lensing contribution.

\item{\it Type Ia Supernova (SN)}

We employ the Union Supernova Ia data (307 SN Ia samples) from The Supernova Cosmology Project \cite{Kowalski2008}.

\item{\it Large Scale Structure (LSS)}

The 2dF Galaxy Redshift Survey (2dFGRS) data \cite{Cole2005} and Sloan Digital Sky Survey (SDSS) Luminous Red Galaxy (LRG) data 
release 4 \cite{Tegmark2006} are utilized.

Note that we have used the likelihood code based on the non-linear
modeling by Tegmark et al.\  \cite{Tegmark2006} (marginalizing the
bias $b$ and the $Q$ parameter). However with a large bump in the linear
power spectrum, this naive treatment may be not sufficient to characterize 
the non-linear response to the feature on small scales.  Ideally, this should be
obtained from N-body simulations, however, such a study is beyond the scope of 
the current work.

There are several other caveats on our results in the high-$k$ regime.  First, we assume 
linear bias for the galaxies, which may not be entirely safe at sufficiently small scales.
Moreover, sharp features in the matter power spectrum can cause sharp features in the bias 
as a function of $k$.

Keeping in mind these caveats our constraints on small scales $k \gsim 0.1\, \mathrm{Mpc}^{-1}$ should be 
taken with a grain of salt and considered as accurate only up to factors order unity.

\item{\it Weak Lensing (WL)}

 Five WL data sets are used in this paper. The effective survey area $A_{\textrm{eff}}$ and galaxy number density $n_{\textrm{eff}}$ of each survey are
 listed in Table~\ref{tblwldata_test_yes}.

\begin{table}[htbp]
{\caption{Weak Lensing Data Sets}\label{tblwldata_test_yes}}
  \begin{center}
  \begin{tabular}{lll} 
  \hline
  \hline
  Data sets& $A_{\textrm{eff}}$  & $n_{\textrm{eff}}$ \\
   & (deg$^2$) & (arcmin$^{-2}$) \\
  \hline
  COSMOS \cite{Massey2007,Lesgourgues2007}  & 1.6 & 40\\
  CFHTLS-wide \cite{Hoekstra2006,Schimd2007} & 22 & 12 \\
  GaBODS \cite{Hoekstra2002a,Hoekstra2002b} & 13 & 12.5 \\
  RCS \cite{Hoekstra2002a,Hoekstra2002b} & 53 & 8 \\
  VIRMOS-DESCART \cite{Van-Waerbeke2005,Schimd2007} & 8.5 &  15 \\
  \hline
  \end{tabular}
  \end{center}
\end{table}

For COSMOS data we use the CosmoMC plug-in written by Julien Lesgourgues \cite{Lesgourgues2007}, modified to do numerical 
marginalization on three nuisance parameters in the original code.

For the other four weak lensing data sets we use the likelihood given by \cite{Benjamin2007}. To calculate the likelihood we have written a CosmoMC 
plug-in code, with simplified marginalization on the parameters of galaxy number density function $n(z)$. More details about this plug-in can be 
found in \cite{Amigo2008}.

As for the LSS data, for small scales $k \gsim 0.1\, \mathrm{Mpc}^{-1}$ there is the caveat that the nonlinear evolution of the power spectrum in the presence of bump-like distortions may not be treated
accurately.
\end{description}


\section{Observational Constraints\label{sec:cons}}

We now present our results for the observational constraints on particle production during inflation, assuming two different scenarios.

\subsection{A Single Burst of Particle Production}\label{subsec:single}

\begin{figure}[tbp]
\begin{center}
\includegraphics[width=3.2in]{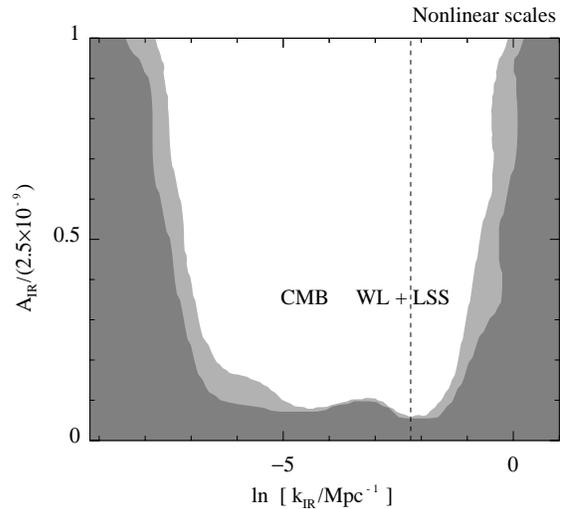}
\caption{Marginalized posterior likelihood contours for the parameters $A_{\mathrm{IR}}$ and $k_{IR}$ (the magnitude and position of the feature, respectively) in the single-bump model. Black and grey regions 
         correspond to parameter values allowed at 95.4\% and 99.7\% confidence levels, respectively.  At small scales, to the right of the dashed vertical line, our results should be taken with a grain of salt since
         the nonlinear evolution of the power spectrum may not be modeled correctly in the presence of bump-like distortions.\label{fig_2d}}
\end{center}
\end{figure}

The minimal scenario to consider is a single burst of particle production during inflation, which corresponds to taking $n=1$ in (\ref{multiple}).
The power spectrum is given by (\ref{param}) with $n=1$ and, with some abuse of notation, we denote $k_1 \equiv k_{\mathrm{IR}}$ and $A_1 \equiv A_{\mathrm{IR}}$. 
The prior we have used for $A_{\text{IR}}$ is $0\le A_{\text{IR}} \le 25\times 10^{-10}$, and for $k_{\text{IR}}$ is $-9.5\le \ln[k/\text{Mpc}^{-1}] \le 1$.  The former condition
ensures that the bump-like feature from IR cascading does not dominate over the observed scale invariant fluctuations while the latter is necessary in order to have the feature in
the observable range of scales.  In Fig.~\ref{fig_2d} we plot the marginalized posterior likelihood for the new parameters $A_{\text{IR}}$ and $k_{\text{IR}}$ describing the
magnitude and location of the bump while in Table \ref{Table_irpool} we give the best fit values for the remaining (vanilla) cosmological parameters.

\begin{table}
\begin{center} 
\caption{Constraints on the standard (``vanilla'') cosmological parameters for the single-bump model.  All errors are 95.4\% confidence level}
\label{Table_irpool}
\begin{tabular}{|c|c|}
\hline
$\Omega_bh^2$ & $0.0227^{+0.0010}_{-0.0010}$ \\
\hline
$\Omega_ch^2$ & $0.1122^{+0.0050}_{-0.0044}$ \\
\hline
$\theta$ & $1.0424^{+0.0042}_{-0.0043}$ \\
\hline
$\tau$ & $0.08^{+0.03}_{-0.03}$ \\
\hline
$n_s$ & $0.956^{+0.024}_{-0.024}$ \\
\hline
$\ln [10^{10}A_s]$ & $3.206^{+0.074}_{-0.068}$ \\
\hline
$A_{SZ}$ & $1.62^{+0.76}_{-0.74}$ \\  
\hline
$\Omega_m$ & $0.264^{+0.026}_{-0.022}$ \\
\hline
$\sigma_8$ & $0.807^{+0.034}_{-0.030}$ \\
\hline
$z_{re}$ & $10.5^{+2.5}_{-2.7}$ \\
\hline
$H_0$ & $71.6^{+2.3}_{-2.4}$ \\
\hline
\end{tabular}
\end{center}  
\end{table}

For very large scales $\lsim \mathrm{Gpc}^{-1}$, the data do not contain much information (due to cosmic variance) and hence the constraint on any modification of the power spectrum
is weak.  In this region the spectral distortion may be larger than $100\%$ of the usual scale invariant fluctuations and couplings $g^2$ order unity are allowed.  For smaller scales
$k \gsim \mathrm{Gpc}^{-1}$ the constraints are stronger and we have, very roughly, $A_{\mathrm{IR}} / A_s \lsim 0.1$ corresponding to $g^2 \lsim 0.01$.  For very small scales, $k \gsim 0.1 \, \mathrm{Mpc}^{-1}$ 
our constraints should be taken with a grain of salt since the nonlinear evolution of the power spectrum may not be modeled correctly in the presence of bump-like distortions.  At small scales nonlinear effects 
tend to wipe out features of this type (see, for example, \cite{Springel:2005nw}) and hence observational constraints for $k \gsim 0.1 \, \mathrm{Mpc}^{-1}$ may be weaker than what is presented in Fig.~\ref{fig_2d}.  
Note that in most of this nonlinear regime we find essentially no constraint on $A_{\mathrm{IR}}$, which is consistent with what would be expected in a more comprehensive treatment.

\begin{figure}[tbp]
\begin{center}
\includegraphics[width=3in]{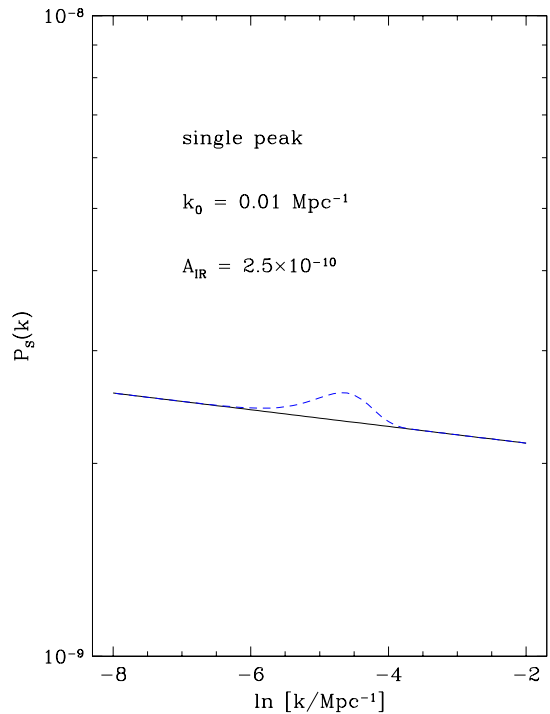}
\includegraphics[width=3in]{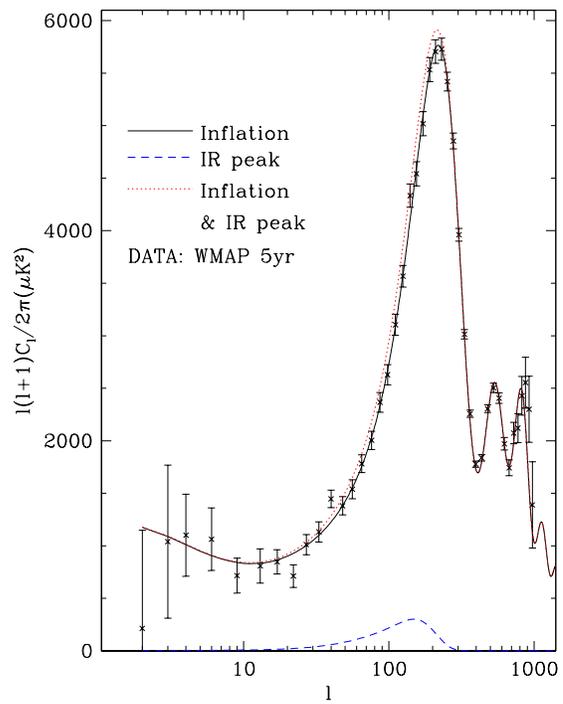}
\caption{The top panel shows a sample bump in the power spectrum with amplitude $A_{\mathrm{IR}} = 2.5\times 10^{-10}$ which corresponds to a coupling $g^2 \sim 0.01$.
The feature is located at $k_{\mathrm{IR}} = 0.01 \, \mathrm{Mpc}^{-1}$.  This example represents a distortion of $\mathcal{O}(10\%)$ as compared to the usual vacuum fluctuations 
and is consistent with the data at $2\sigma$.  The bottom panel shows the CMB angular TT power spectrum for this example, illustrating that the distortion shows up mostly in the first peak.}\label{Fig:sample}
\end{center}
\end{figure}

The IR cascading bump in the primordial power spectrum will be accompanied by a corresponding nongaussian feature in the bispectrum 
\cite{ir,inprog2}.  From the perspective of potentially observing this signal it is most interesting if this feature is located on scales
probed by CMB experiments.  (There is also the fascinating possibility that the nongaussianity from IR cascading could show up in the large scale
structure as in \cite{dalal,shandera,mcdonald,afshordi}.  We leave a detailed discussion to future studies.)
To get some intuition into what kinds of features in the CMB scales are still allowed by the data we focus on an example with
$A_{\text{IR}}=2.5\times 10^{-10}$ which, using (\ref{Ag}), corresponds to a reasonable coupling value $g^2 \sim 0.01$.  We take the bump to be
located at $k_{\mathrm{IR}} = 0.01\, \mathrm{Mpc}^{-1}$ and fix the remaining model parameters to $A_s=2.44\times 10^{-9}$, $n_s=0.97$ (which are compatible
with the usual values).  This sample bump in the power spectrum is illustrated in the top panel of Fig.~\ref{Fig:sample} and is consistent with the data at $2\sigma$.
In the bottom panel of Fig.~\ref{Fig:sample} we plot the associated angular CMB TT spectrum.  This example represents a surprisingly large spectral 
distortion: the total power in the feature as compared to the scale invariant vacuum fluctuations is $P_{\mathrm{bump}} / P_{\mathrm{vac}} \sim 0.1$, 
evaluated at the peak of the bump.  In \cite{inprog2} we discuss the nongaussianity associated with this feature.

\subsection{Multiple Bursts of Particle Production\label{subsec:multiple}}

Next, we consider a slightly more complicated scenario: multiple bursts of particle production leading many localized features
in the power spectrum.  For simplicity we assume that all bumps have the same magnitude $A_i \equiv A_{\mathrm{IR}}$ and we further
suppose a fixed number of $e$-foldings $\delta N$ between each burst of particle production.  This implies that the location
of the $i$-th bump will be given by $k_i = e^{(i-1)\Delta} k_1$ where $\Delta$ is a model parameter controlling the density of features.
We take the number of bursts, $n$, to be sufficiently large that the series of features extends over the whole observable range.  In
the next section we will see that these assumptions are not restrictive and that many well-motivated models are consistent with this simple set-up.  

\begin{figure}[tbp]
\begin{center}
\includegraphics[width=3.2in]{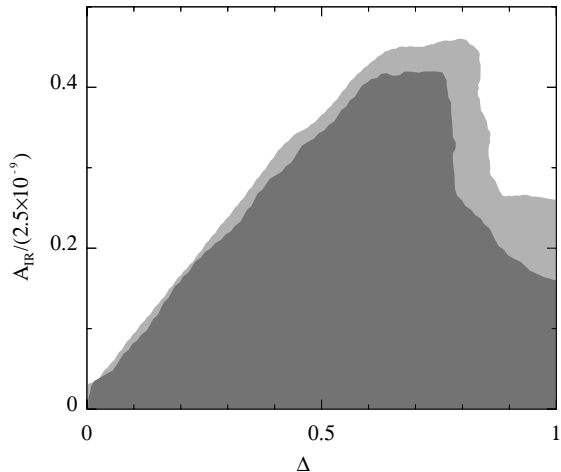}
\caption{Marginalized posterior likelihood contours for the parameters $A_{\mathrm{IR}}$ and $\Delta$ (the feature amplitude and spacing, respectively) of the multiple-bump model. Black and grey regions 
               correspond to values allowed at 95.4\% and 99.7\% confidence levels, respectively. } \label{fig_2d_multiple}
\end{center}
\end{figure}

Our multi-bump model, then, has three parameters: $A_{\mathrm{IR}}$, $k_1$ and $\Delta$.  We take the prior on the amplitude to be $A_{\mathrm{IR}} \leq 25\times 10^{-10}$ as
in section \ref{subsec:single}.  If the features are very widely spaced, $\Delta \gsim 1$, then the constraint on each bump will obviously be identical to the
results for the single-bump case presented in the section \ref{subsec:single}.  Hence the most interesting case to consider is $\Delta \lsim 1$ so that the bumps are partially overlapping.
Our prior for the density of features is therefore $0 \leq \Delta \leq 1$.  Finally, the location of the first bump
will be a historical accident in realistic models, hence we marginalize over all possible values of $k_1$ and present our constraints and 2-d likelihood plots in the space
of $A_{\mathrm{IR}}$ and $\Delta$.  This marginalized likelihood plot is presented in Fig.~\ref{fig_2d_multiple}.  In table \ref{Table_mp_close} we present the best-fit values for the vanilla cosmological
parameters.

\begin{table}
\begin{center} 
\caption{constraints on the standard (``vanilla'') cosmological parameters for the multiple-bump model.  All error bars are 95.4\% confidence level.}
\label{Table_mp_close}
\begin{tabular}{|c|c|}
\hline
$\Omega_bh^2$ & $0.0227^{+0.0009}_{-0.0009}$ \\
\hline
$\Omega_ch^2$ & $0.1126^{+0.0049}_{-0.0044}$ \\
\hline
$\theta$ & $1.0424^{+0.0039}_{-0.0043}$ \\
\hline
$\tau$ & $0.078^{+0.031}_{-0.026}$ \\
\hline
$n_s$ & $0.93^{+0.04}_{-0.17}$ \\
\hline
$\ln [10^{10}A_s]$ & $2.8^{+0.4}_{-0.9}$ \\
\hline
$A_{SZ}$ & $1.60^{+0.77}_{-0.76}$ \\
\hline
$\Omega_m$ & $0.265^{+0.026}_{-0.021}$ \\
\hline
$\sigma_8$ & $0.807^{+0.034}_{-0.030}$ \\
\hline
$z_{re}$ & $10.3^{+2.6}_{-2.5}$ \\
\hline
$H_0$ & $71.4^{+2.2}_{-2.4}$ \\
\hline
\end{tabular}
\end{center}  
\end{table}

From the likelihood plot, Fig.~\ref{fig_2d_multiple}, there is evidently a preferred value of the feature spacing, roughly $\Delta \sim 0.75$, for which the constraints are weakest.  This can be understood
as follows.  For very high density $\Delta \rightarrow 0$ the localized features from IR cascading smear together and the total power spectrum (\ref{param}) is $P(k) \sim A_s (k / k_0)^{n_s - 1} + C$ where the 
size of the constant deformation scales linearly with the density of features: $C \propto \Delta^{-1}$.  Therefore, the upper bound on the amplitude $A_{\mathrm{IR}}$ should scale linearly with $\Delta$.  Indeed, this linear
trend is very evident from Fig.~\ref{fig_2d_multiple} in the small-$\Delta$ regime.  This linear behaviour must break down at some point since as the features become infinitely widely spaced the constraint
on $A_{\mathrm{IR}}$ must go to zero.  This explains the bump in the likelihood plot, Fig.~\ref{fig_2d_multiple}, near $\Delta \sim 0.75$.

In passing, notice that the behaviour $P(k) \sim A_s (k / k_0)^{n_s - 1} + C$ for $\Delta \ll 1$ also explains why the best-fit $A_s$ in table \ref{Table_mp_close} is somewhat lower than the standard value and why the spectral tilt
$n_s-1$ is somewhat more red.

\begin{figure}[tbp]
\begin{center}
\includegraphics[width=3in]{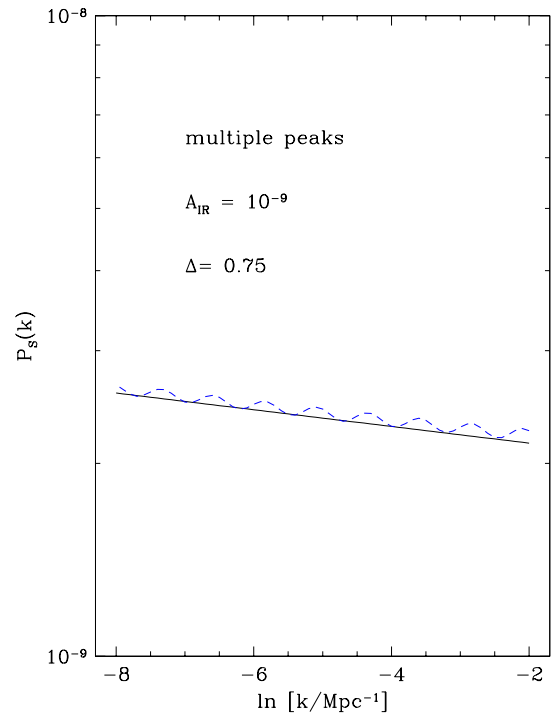}
\includegraphics[width=3in]{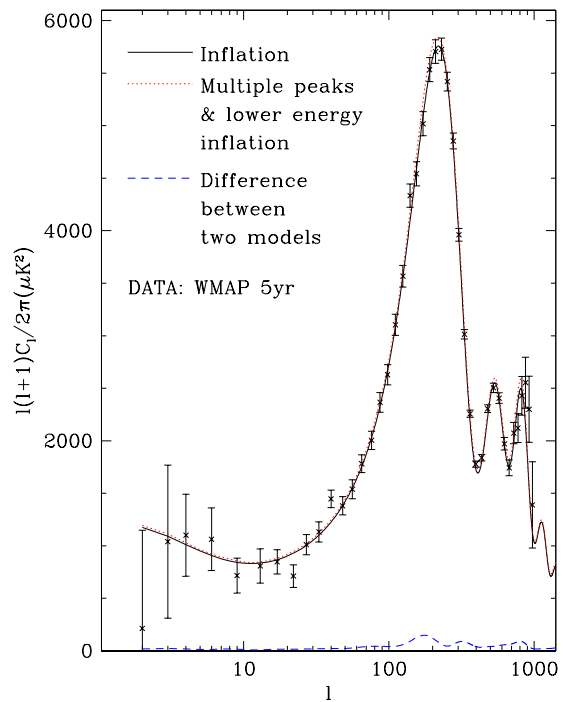}
\caption{The top panel shows a sample multiple-bump distortion with amplitude $A_{\mathrm{IR}} = 1\times 10^{-9}$ which corresponds to $g^2 \sim 0.02$.  The feature spacing is $\Delta = 0.75$.
We take the vanilla parameters to be $A_s = 1.04\times 10^{-9}$, $n_s = 0.93$ so that the scale of inflation is slightly lower than in the standard scenario and the spectral tilt is slightly redder.
The bottom panel shows the CMB angular TT power spectrum for this example.}\label{Fig:sample2}
\end{center}
\end{figure}

To get some intuition for the kinds of multi-bump distortions that are allowed by the data, we consider an example with $A_{\mathrm{IR}} = 1 \times 10^{-9}$, $\Delta = 0.75$ and fix the vanilla parameters to 
$A_s=1.04\times10^{-9}$, $n_s=0.93$.  This choice of parameters is consistent with the data at $2\sigma$ and corresponds to a reasonable coupling $g^2 \sim 0.02$.  In Fig.~\ref{Fig:sample2} we plot 
the primordial power spectrum $P(k)$ and also the CMB TT angular power spectrum for this example.

\section{Particle Physics Models\label{sec:particle}}

From the low energy perspective one expects interactions of the type (\ref{int}) to be rather generic, hence particle production during
inflation may be expected in a wide variety of models.  In this section we consider some explicit examples
in string theory and SUSY in order to show how such models may be obtained microscopically and also to provide
the proof of concept that realistic models do exist where $\phi_i$ are in the observable range.

\subsection{Open String Inflation Models}

String theory inflation models may be divided into two classes depending on the origin of the inflaton: closed string models and open string models.
In the former case the inflaton is typically a geometrical modulus associated with the compactification manifold (examples include racetrack
inflation \cite{racetrack}, K\"ahler modulus inflation \cite{CQ} and Roulette inflation \cite{roulette}).  In the latter case the 
inflaton is typically the position modulus of some mobile D-brane\footnote{One notable exception is inflation driven by the open string tachyon, for example nonlocal string field theory models \cite{nonlocal}.}
living in the compactification manifold (examples include brane inflation \cite{brane} such as the warped KKLMMT model \cite{KKLMMT}, 
D3/D7 inflation \cite{D3D7} and DBI inflation \cite{DBI}).  In open string inflation models there may be, in addition to the mobile inflationary brane, 
some additional ``spectator'' branes.  If the mobile brane collides with any spectator brane during inflation then some of the stretched string states between these branes will become massless
at the moment when the branes are coincident \cite{beauty,trapped}, precisely mimicking the interaction (\ref{int}).  Thus, we expect particle 
production, IR cascading and the bump-like features described above to be a reasonably generic prediction of open string inflation.

\subsection{String Monodromy Models}

A concrete example of the heuristic scenario discussed in the last subsection is provided by the brane monodromy
and axion monodromy string theory inflation models proposed in \cite{monodromy,monodromy2,monodromy3}.  
In the original brane monodromy model \cite{monodromy} one considers type IIA string theory compactified on a nil manifold that is the
product of two twisted tori.  The metric on each of these twisted tori has the form
\begin{equation}
  \frac{ds^2}{\alpha'} = L_{u_1}^2 du_1^2 +  L_{u_2}^2 du_2^2 + L_x^2 (dx' + M u_1 du_2)^2
\end{equation}
where $x' = x- \frac{M}{2} u_1 u_2$ and $M$ is an integer flux number.  The dimensionless constants $L_{u_1}$, $L_{u_2}$ and $L_x$ 
determine the size of the compactification.

Inflation is realized by the motion of a D4-brane along the direction $u_1$ of the internal manifold.
The D4 spans our large 3-dimensions and wraps a 1-cycle along the direction $u_2$ of the internal space.  The size of this 1-cycle, in string units, is given by 
\begin{equation}
\label{L}
  L = \sqrt{L_{u_2}^2 + L_x^2 M^2 u_1^2}
\end{equation}
Hence, the brane prefers to minimize its world-volume by moving to the location $u_1 = 0$ where this 1-cycle has minimal size.  This preference gives a potential to 
the D4-brane position which goes like $V \sim u_1$ in the large $u_1$ regime that is relevant for large field inflation.  

In \cite{trapped} it was shown that this scenario allows for the inclusion of a number of spectator branes stabilized at positions $u_{1} = i / M$ (with $i$ integer) 
along the inflationary trajectory.  As the mobile inflationary D4 rolls through these points particles (strings) will be produced and the resulting distribution of
features will look nearly identical to the simple multi-bump scenario studied in section \ref{subsec:multiple}.  To see this, let us now determine the distribution of 
features that is predicted from brane monodromy inflation.  The canonical inflaton $\phi$ can be related to the position of the mobile D4 as
\begin{equation}
\label{canonical}
  \phi = B\, u_1^{1/p}
\end{equation}
where $B$ is a constant with dimensions of mass that depends on model parameters.  Hence, the effective potential during inflation has the power-law form
\begin{equation}
\label{monodromy_pot}
  V(\phi) = \mu^{4-p} \phi^p
\end{equation}
For the simplest scenario described above one has $p = 2/3$.  However, the formulas (\ref{canonical},\ref{monodromy_pot}) still hold 
for the variant considered in \cite{monodromy} with $p = 2/5$ as long as one replaces $u_1$ by a more complicated linear combination of coordinates.  These relations also hold for axion
monodromy models \cite{monodromy2} with $p=1$ and $u_1$ replaced by the axion, $c$, arising from a 2-form RR potential $C^{(2)}$ integrated over a 2-cycle $\Sigma_2$.  For \emph{all} models of the form (\ref{monodromy_pot})
the number of $e$-foldings $N$ from $\phi = \phi(N)$ to the end of inflation (which occurs at $\phi = p /\sqrt{2}$ when the slow roll parameter $\epsilon(\phi)= 1$) is given by
\begin{eqnarray}
  N &=& \frac{1}{2p} \frac{\phi^2(N)}{M_p^2} - \frac{p}{4} \nonumber \\
    &=& \frac{1}{2p} \frac{B^2}{M_p^2} u_1^{2/p} - \frac{p}{4} \label{monodromy_N}
\end{eqnarray}
Since the spectator branes are located at $u_1 = i / M$ the bursts of particle production must occur at times $N = N_i$ where
\begin{equation}
  N_i = \frac{1}{2p} \frac{B^2}{M_p^2} \left(\frac{i}{M}\right)^{2/p} - \frac{p}{4}
\end{equation}
The location $k=k_i$ of the $i$-th feature is defined, roughly, by the scale leaving the horizon at the moment $N=N_i$.  Hence, the distribution of features for brane/axion monodromy models is given by
\begin{equation}
\label{monodromy_distribution}
  \ln \left[\frac{k_i}{H} \right] \cong \tilde{B}^2 i^{2/p} - \frac{p}{4}
\end{equation}
with $p = 2/3$ or $p=2/5$ for brane monodromy and $p=1$ for axion monodromy.  In (\ref{monodromy_distribution}) the dimensionless number $\tilde{B}$ depends on model parameters.

Although the distribution of features (\ref{monodromy_distribution}) is not exactly the same as the evenly space distribution considered subsection \ref{subsec:multiple}, the two are essentially indistinguishable
over the range of scales which are probed observationally (corresponding to roughly 10 $e$-foldings of inflation).  The reason for this is simple: the inflaton is nearly constant during the first 10 $e$-foldings of inflation
and hence $\delta N \sim \delta \phi \sim \delta u_1$ within the observable region.  It follows that $k_i \cong e^{(i-1)\Delta} k_1$ to very good approximation for a huge class of models.  To see this more concretely in
the case at hand, let us compute $dN / du_1$  from (\ref{monodromy_N}).  It is straightforward to show that
\begin{equation}
  \frac{dN}{du_1} = \frac{1}{p^p} \frac{1}{\left[2\epsilon(\phi)\right]^{1-p/2}} \left( \frac{B}{M_p} \right)^{p}
\end{equation}
where
\begin{equation}
  \epsilon(\phi) \equiv \frac{M_p^2}{2}\left(\frac{V'}{V}\right)^2 = \frac{p^2}{2} \left(\frac{M_p}{\phi}\right)^2
\end{equation}
is the usual slow roll parameter.  Observational constraints on the running of the spectral index imply that $\epsilon(\phi)$ cannot change much over the observable 10 $e$-foldings of inflation.
Since $dN/du_1 \cong \mathrm{const}$ to very high accuracy it follows trivially that $N = N(u_1)$ is very close to linear and $k_i \cong e^{(i-1)\Delta} k_1$ as desired.

In the context of axion monodromy inflation models \cite{monodromy2} the multiple bump features discussed here will be complimentary to the oscillatory features described in \cite{monodromy3}
which result from the sinusoidal modulation of the inflaton potential by instanton effects.  If the bursts of particle production are sufficiently densely spaced, then signal from IR cascading may appear
oscillatory, however, it differs from the effect discussed in \cite{monodromy3} in both physical origin and also in functional form.

Let us now estimate the effective value of the couplings $g^2_i$ appearing in the prototype interaction (\ref{multiple}) that are predicted from the simplest brane monodromy model.  A complete calculation would involve
dimensionally reducing the DBI action describing the brane motion and requires knowledge of the full 10-dimensional geometry with the various embedded branes.  For our purposes, however, a simple heuristic estimate
for the collision of two D4-branes will suffice.  When $N$ D-branes become coincident the symmetry is enhanced from $U(1)^N$ to a $U(N)$ Yang Mill gauge theory.  The gauge coupling for this Yang Mills theory is given
by
\begin{equation}
  g_{\mathrm{YM}}^2 = \frac{g_s (2\pi)^2}{L}
\end{equation}
where $L$ is the volume of the 1-cycle that the D4 branes wrap and is given by (\ref{L}).  If the inflationary brane is at position $u_1$ and the $i$-th spectator brane is at $u_{1,i}$ then the distance between
the two branes is given by
\begin{equation}
  d^2 = \alpha'\, L_{u_1}^2 (u_1 - u_{1,i})^2
\end{equation}
The mass of the gauge bosons corresponding to the enhanced symmetry is
\begin{equation}
\label{boson_mass}
  M^2_i = g_{\mathrm{YM}}^2 \frac{d^2}{(2\pi)^2 (\alpha')^2} = \frac{g_s L_{u_1}^2\, (u_1 - u_{1,i})^2}{\alpha' \sqrt{L_{u_2}^2 + L_x^2 M^2 u_1^2}}
\end{equation}
To put this in the prototype form $M^2_i = g_i^2 (\phi - \phi_i)^2$ we must first convert to the canonical variable $\phi$ using the formula (\ref{canonical}) with $p = 2/3$ and 
\begin{equation}
  B = \frac{M^{1/2}}{6\pi^2}\frac{L_{u_1}L_x^{1/2}}{\sqrt{g_s \alpha'}}
\end{equation}
Next, we must Taylor expand the resulting equation about the minimum $\phi = \phi_i$.  We find
\begin{eqnarray}
  M_i^2 &\cong& g_i^2 (\phi- \phi_i)^2 + \cdots \\
  g_i^2 &=& \frac{16 g_s^2 \pi^4}{M L_x u_{1,i}} \frac{1}{\sqrt{L_{u_2}^2 + L_x^2 M^2 u_{1,i}^2}} \nonumber \\
        &=&  \frac{16 g_s^2 \pi^4}{ L_x i} \frac{1}{\sqrt{L_{u_2}^2 + L_x^2 i^2}} \label{g_monodromy}
\end{eqnarray}
where on the second line of (\ref{g_monodromy}) we have used the fact that $u_{1,i} = i / M$ (with $i$ integer) in the simplest models.  We see that the effective couplings $g_i^2$ become larger as the D4 unwinds during inflation.  
(The apparent divergence for $u_{1,i} = 0$ in the formula (\ref{g_monodromy}) is an artifact of the fact that the relation (\ref{canonical}) is not valid at small values of $u_1$.  This will not concern us here since inflation 
has already terminated at the point that our formulas break down.)

To compute the amplitude of the bump-like feature produced by brane monodromy inflation we should take into account also combinatorial factors.  When two branes become coincident the symmetry is enhanced from $U(1)^2$ to $U(2)$ 
so there are $2^2 - 2 = 2$ additional massless spin-$1$ fields appearing at the brane collision.  Thus, using equation (\ref{Ag}), the amplitude of the feature that will be imprinted in the CMB is
\begin{equation}
\label{Aeff}
  A_{i,\mathrm{eff}} = 2\times (2^2 - 2) \times \left[ 1.01\cdot10^{-6}\cdot g_i^{15/4} \right]
\end{equation}
where the extra factor of $2$ counts the polarizations of the massless spin-$1$ fields.  This combinatorial enhancement can be much larger if the inflationary brane collides with a \emph{stack} of spectators.

The above discussion is predicated on the assumption that the original brane monodromy set-up \cite{monodromy} is supplemented by additional spectator branes.  This may seem like an unnecessary contrivance,
however, in order for this model to reheat successfully it may be \emph{necessary} to include spectator branes.  For example, with the reheating mechanism proposed in \cite{anke} semi-realistic 
particle phenomenology can be obtained by confining the standard model (SM) to a D6 brane which wraps the compact space.  In order to reheat into SM degrees of freedom we orient this brane so that its world-volume is 
parallel to the mobile (inflationary) D4.  In this case the end of inflation involves multiple oscillations of the D4 about the minimum of its potential.  At each oscillation the D4 collides
with the D6 and SM particles are produced by parametric resonance preheating \cite{kls1,kls2}.  However, due to the periodic structure of the compactification, D4/D6 collisions will necessarily occur also \emph{during} 
inflation, leading to IR cascading features in the CMB.  

The timing of these D4/D6 collisions was computed in \cite{anke} for the minimal $p = 2/3$ brane monodromy model, assuming the same choices of parameters used in 
\cite{monodromy}.  For this particular case there is only one collision (and hence one feature) during the first 10 $e$-foldings of inflation and the phenomenology is essentially the same as that considered in subsection 
\ref{subsec:single}.  What is the amplitude of this feature?  Assuming, again, the parameters employed in \cite{monodromy} and noting that the first collision takes place at $i = 13$ \cite{anke} equation (\ref{g_monodromy})
gives $g_1^2 \cong 0.001$.  From (\ref{Aeff}) we find the effective amplitude of the feature to be $A_{1,\mathrm{eff}} / A_s \cong 0.01$.  This value is well within the observational bounds derived in subsection \ref{subsec:single} 

We stress that the conclusions in the previous paragraph apply \emph{only} for the particular choice of model parameters employed in \cite{monodromy}.  There exist other consistent parameter choices for which the simplest brane monodromy 
model predicts a much higher density of features with much larger amplitude.

Note that both brane and axion monodromy models may be used to realize trapped inflation \cite{trapped}.  Here we are restricting ourselves to 
the large-field regime where the potential $V = \mu^{4-p}\phi^p$ is flat enough to drive inflation without the need for trapping effects.
For a given choice of parameters one should verify that this classical potential dominates over the quantum corrections from particle production.

\subsection{A Supersymmetric Model}

Another microscopic realization of multiple bursts of particle production and IR cascading during inflation which does not
rely on string theory can be obtained from the so-called ``distributed mass'' model derived in \cite{berrera} with warm inflation \cite{warm}
in mind, however, the theory works equally well for our scenario.  This model is based
on $\mathcal{N} =1$ global SUSY and allows for the inclusion of multiple points along the inflationary trajectory where both scalar degrees of 
freedom and also their associated fermion superparteners become massless.  The distribution of features in this set-up is essentially arbitrary.

\section{Conclusions\label{sec:concl}}

In this paper we have studied the observational constraints on models with particle production during inflation.  We have focused on the simple
prototype model (\ref{int}) for each burst of particle production, 
however, we expect that our qualitative results will apply also to more complicated models (for example with gauged
interactions or fermion iso-inflaton fields) and perhaps also to the case of inflationary phase transitions.  We find no evidence for a detection
of the features associated with particle production and IR cascading, however, it is interesting to note that rather large localized features are still
compatible with the data.   Our results differ significantly from previous studies as a result of a more realistic treatment of the 
cosmological perturbations in models with particle production.  The bounds we have derived on $g^2$ will play a crucial role in assessing the detectability 
of the nongaussianity produced by particle production and IR cascading.    

We have also discussed the implications of our results for popular brane/axion monodromy string theory inflation models.  Successful reheating in these constructions
may require the inclusion of spectator branes which collide with the mobile D4-brane during inflation and hence we expect CMB features to be a fairly generic prediction.  
We have shown that brane/axion monodromy models predict a distribution of bump-like features which are evenly spaced in $\ln k$ over the observable range
of scales.  In the case of axion monodromy this multiple-bump spectral distortion is complimentary to the oscillatory features discussed in \cite{monodromy3}.  We have
also estimated the magnitude of these bump-like features in terms of model parameters.

One motivation for the present study was to determine the extent to which microscopically realistic models such as (\ref{int}) can reproduce
the localized ``glitches'' in the power spectrum that have been detected (albeit with marginal significance) by several previous studies.  These
anomalies can be classified as follows:
\begin{enumerate}
  \item \emph{Localized power excesses}:  

  \hspace{2mm} In both \cite{chung2} and \cite{hoi} power spectra with localized spikes were studied and in both cases 
  marginal evidence was found for a detection of such features.  In \cite{yokoyama2} a non-parametric reconstruction of the power spectrum was performed and
  the result is marginally consistent with a power law everywhere, however, several localized spikes are evident in the reconstruction.

  \hspace{2mm} Localized excesses are naturally obtained in our model (\ref{int}).  Sadly, however, we did
  not find that our model fits the data significantly better than the simplest slow roll inflation scenario.  This does not necessarily imply a
  disagreement with \cite{chung2,hoi} since we use a different shaped feature and different data sets. (Indeed, when the authors of \cite{hoi} repeat
  their analysis using the WMAP 5-year data they do not obtain a detection \cite{jim_feat}, consistent with our findings.)

  \item \emph{Localized power deficits}:  

  \hspace{2mm} In \cite{features2} the Richardson-Lucy deconvolution algorithm was used to perform a non-parametric 
  reconstruction of the primordial power spectrum which displayed a prominent IR cut-off near the horizon.  In \cite{contaldi} a similar analysis
  was performed and the reconstructed power spectrum displays a localized dip in power near $k \sim 0.002 \, \mathrm{Mpc}^{-1}$.

  \hspace{2mm}  Localized deficits \emph{can}
  be produced by our model (\ref{multiple}) but only in a rather contrived way.  Hence, we have not focused on such features in section \ref{sec:cons}.

  \item \emph{Damped oscillations}:  

  \hspace{2mm} In \cite{adams,morefeatures1,morefeatures2,gobump} power spectra with superimposed ringing patterns
  were studied.  Such features provide a marginally improved fit over the simplest power-law model.

  \hspace{2mm} As we have discussed in the introduction, damped oscillatory ``ringing'' features are not predicted by inflationary particle production.
  Nor is it clear if such features are predicted by models with phase transitions.  (Of course damped oscillations \emph{can} be obtained
  from a toy model with a step in $V(\phi)$.  However, it may be difficult to obtain such a potential from realistic micro-physics; generically 
  one expects that any sharp features in $V(\phi)$ will be 
  smoothed out by quantum corrections.)
\end{enumerate}

Finally, let us note that features of the type studied here will lead to other observables beyond the distortion of the primordial
power spectrum.  In particular, bumps in $P(k)$ will lead to features in the tensor spectrum (resulting from
the sourcing of gravitational waves by scalar fluctuations at second order in perturbation theory) and also, possibly, black hole production.
In \cite{yokoyama4,yokoyama5} these effects were estimated assuming a power spectrum which is qualitatively similar to ours.  
As discussed in \cite{ir}, inflationary particle production will also lead to potentially large localized nongaussian features in the bispectrum
(and higher order statistics) of the cosmological fluctuations.  These nongaussianities will be discussed in detail in an upcoming work \cite{inprog2}.

\bigskip{\noindent {\bf Acknowledgments}}  

This work was supported in part by NSERC.  We are grateful to J.~Cline, N.~Dalal, H.~Firouzjahi, D.~Green, L.~Hoi, L.~Kofman, P.~McDonald and G.~D.~Moore for
helpful comments and discussions.


\end{document}